\documentclass[vecphys]{svmult}

\usepackage{amsfonts,amsmath,amssymb,amscd,mathrsfs,cite}
\usepackage{makeidx}         
\usepackage{graphicx}        
\usepackage{multicol}        
\usepackage[bottom]{footmisc}

\makeindex             

\def\bea{\begin{eqnarray}}
\def\eea{\end{eqnarray}}
\def\be{\begin{equation}}
\def\ee{\end{equation}}

\newtheorem{conjec}{Conjecture}
\def\gierN{\mathcal{N}}
\def\gierP{\mathcal{P}}
\def\kett#1{|#1\rangle}
\def\sstyle{\scriptstyle}
\def\gierbinom#1#2{\left(\begin{array}{@{}c@{}} #1 \\ #2 \end{array}\right)}
\DeclareMathOperator{\gierd}{d}
\DeclareMathOperator{\giere}{e}

\begin{document}

\title*{Fully packed loop models on finite geometries}

\author{Jan de Gier}
\institute{Department of Mathematics and Statistics, The University of Melbourne, Victoria 3010, Australia, \texttt{degier@ms.unimelb.edu.au}}

\maketitle

\section{Fully packed loop models on the square lattice}
\begin{figure}[h]
\centerline{
\begin{picture}(200,200)
\put(0,0){\includegraphics[width=200pt]{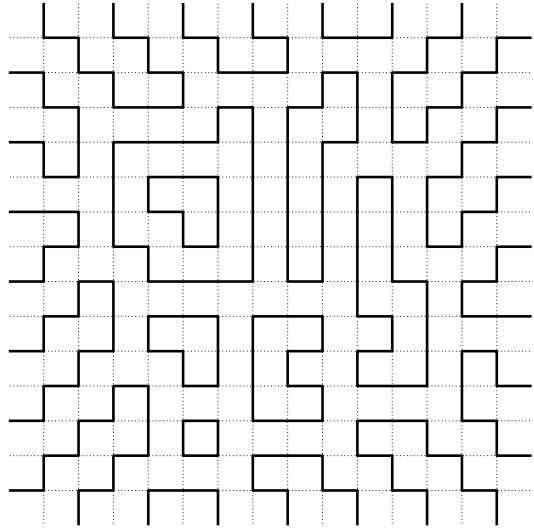}}
\end{picture}}
\caption{Fully packed loops inside a square with alternating boundary condition.}
\label{fig:FPLdef}
\end{figure}

\index{fully packed loops}
A fully packed loop (FPL) model on the square lattice is the statistical ensemble of all loop 
\index{loop configurations}
configurations, where loops are drawn on the bonds of the lattice, and each loop visits every site once \cite{BatchBNY96,DeiContN}. On finite geometries, loops either connect external terminals on the boundary, or form closed circuits, see for example Figure~\ref{fig:FPLdef}. In this chapter we shall be mainly concerned with FPL models on squares and rectangles with an alternating boundary condition 
\index{alternating boundary condition}
where every other boundary terminal is covered by a loop segment, see Figure~\ref{fig:FPLdef}. 

\index{closely packed polygons} 
An FPL model thus describes the statistics of closely packed polygons on a finite geometry. Polygons 
\index{punctures}
may be nested, corresponding to punctures studied in Chapter 8. FPL models can be generalised to include weights. In particular we will study FPL models where a weight $\tau$ is given to each straight local loop segment. The partition function of an FPL model on various geometries can be computed 
\index{partition function}
exactly using its relation to the solvable six-vertex lattice model. It is well known that the model 
\index{six-vertex model} 
undergoes a bulk phase transition at $\tau=2$. \index{phase transition!bulk}

We furthermore study nests of polygons connected to the boundary. In the case of FPL models with mirror or rotational symmetry, the probability distribution function of such nests is known analytically, albeit conjecturally. FPL models undergo another phase transition as a function of the boundary nest fugacity. At criticality, we derive a scaling form for the nest distribution function which displays an 
\index{nest!generating function}
unusual non-Gaussian cubic exponential behaviour.

The purpose of this chapter is to collect and discuss known results for FPL models which may be relevant to polygon models. For that reason we have not put an emphasis on derivations, many of which are well-documented in the existing literature, but rather on interpretations of results.

\subsection{Bijection with the six-vertex model, alternating-sign matrices and height configurations}
\begin{figure}[h]
\centerline{
\begin{picture}(230,106)
\put(30,0){\includegraphics[width=200pt]{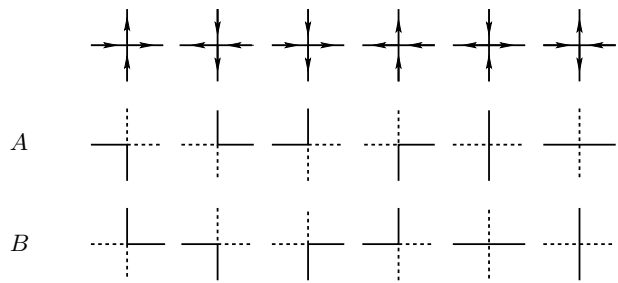}}
\put(0,50){$A$}
\put(0,12){$B$}
\end{picture}}
\caption{Bijection between six-vertex and FPL vertices. The correspondence is different on the two sublattices $A$ and $B$.}
\label{fig:FPLex}
\end{figure}

\index{alternating-sign matrices} \index{height configurations}
There is a well-known one-to-one correspondence between FPL, six-vertex and alternating-sign configurations \cite{RobbinsR86,ElkiesKLP92}. In the six-vertex model, to each bond of the square lattice is associated an arrow, such that at each vertex there are two in- and two out-pointing arrows, see e.g. \cite{Baxter82}. There are six local vertex configurations which are given in the top row of Figure~\ref{fig:FPLex}. The six-vertex and FPL configurations are related in the following way. The square lattice is divided into two sublattices, even ($A$) and odd ($B$). For each arrow configuration we draw only those bonds on which the arrow points to the even sublattice. 
\begin{figure}[h]
\centerline{
\begin{picture}(200,80)
\put(0,0){\includegraphics[width=80pt]{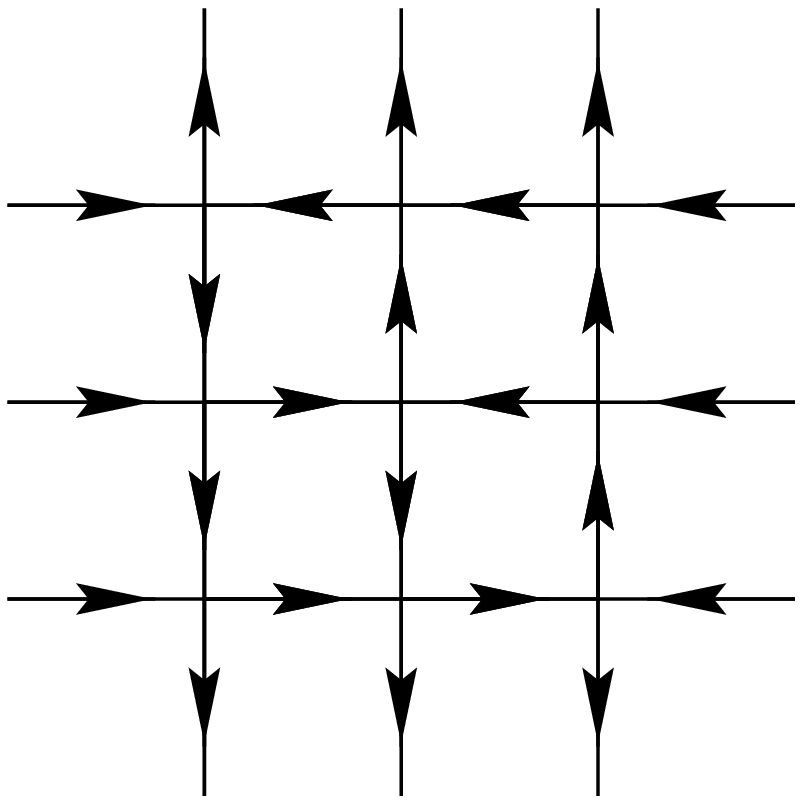}}
\put(120,0){\includegraphics[width=80pt]{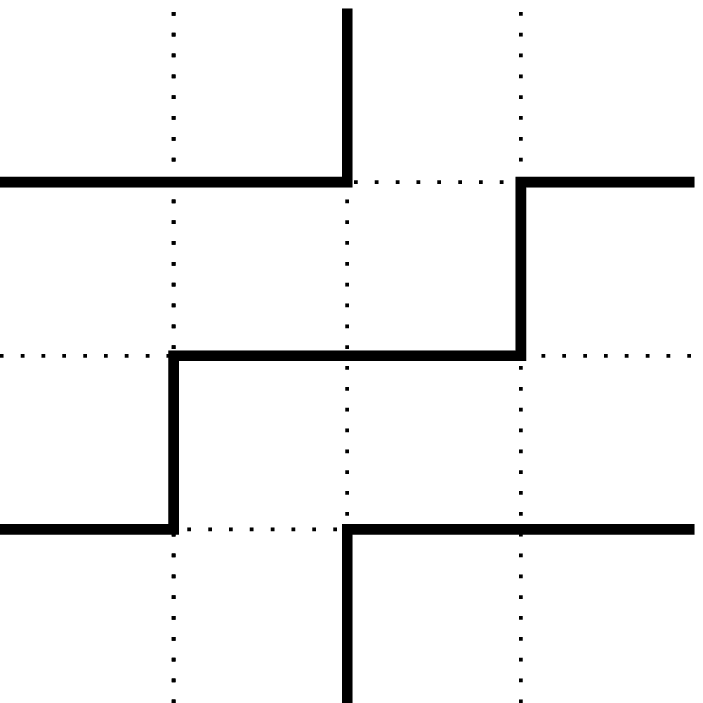}}
\end{picture}}
\caption{An equivalent six-vertex and fully packed loop configuration.}
\label{fig:FPL_6V}
\end{figure}
If we choose the vertex in the upper left corner to belong to the even sublattice, the six-vertex and FPL configuration in Figure~\ref{fig:FPL_6V} are equivalent, as can be seen from the correspondence in Figure~\ref{fig:FPLex}.

\begin{figure}[h]
\centerline{
\begin{picture}(300,53)
\put(0,10){\includegraphics[width=300pt]{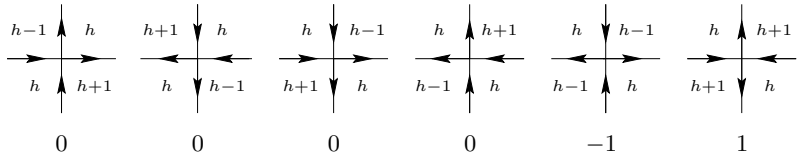}}
\put(19,-4){$0$}\put(70.5,-4){$0$}\put(122,-4){$0$}
\put(173.5,-4){$0$}\put(219.5,-4){$-1$}\put(276.5,-4){$1$}
\put(2,40){$\sstyle h-1$}  
\put(30,40){$\sstyle h$} 
\put(9,19){$\sstyle h$} 
\put(27,19){$\sstyle h+1$}
\put(52,40){$\sstyle h+1$}  
\put(80,40){$\sstyle h$} 
\put(59,19){$\sstyle h$} 
\put(77,19){$\sstyle h-1$}
\put(112,40){$\sstyle h$}  
\put(130,40){$\sstyle h-1$} 
\put(105,19){$\sstyle h+1$} 
\put(133,19){$\sstyle h$}
\put(162,40){$\sstyle h$}  
\put(180,40){$\sstyle h+1$} 
\put(155,19){$\sstyle h-1$} 
\put(183,19){$\sstyle h$}
\put(214,40){$\sstyle h$}  
\put(232,40){$\sstyle h-1$} 
\put(207,19){$\sstyle h-1$} 
\put(235,19){$\sstyle h$}
\put(266,40){$\sstyle h$}  
\put(284,40){$\sstyle h+1$} 
\put(259,19){$\sstyle h+1$} 
\put(287,19){$\sstyle h$}
\end{picture}}
\caption{The six vertices and their corresponding heights and ASM
entries.}
\label{fig:6V}
\end{figure}

\begin{figure}[hfoll]
\centerline{
\begin{picture}(275,115)
\put(0,0){\includegraphics[width=115pt]{Ice-example.eps}}
\put(14,14){$\sstyle 3$}\put(14,42){$\sstyle 2$}\put(14,70){$\sstyle
1$}\put(14,98){$\sstyle 0$}
\put(42,14){$\sstyle 2$}\put(42,42){$\sstyle 1$}\put(42,70){$\sstyle
0$}\put(42,98){$\sstyle 1$}
\put(70,14){$\sstyle 1$}\put(70,42){$\sstyle 0$}\put(70,70){$\sstyle
1$}\put(70,98){$\sstyle 2$}
\put(98,14){$\sstyle 0$}\put(98,42){$\sstyle 1$}\put(98,70){$\sstyle
2$}\put(98,98){$\sstyle 3$}
\put(160,-4){\includegraphics[width=115pt]{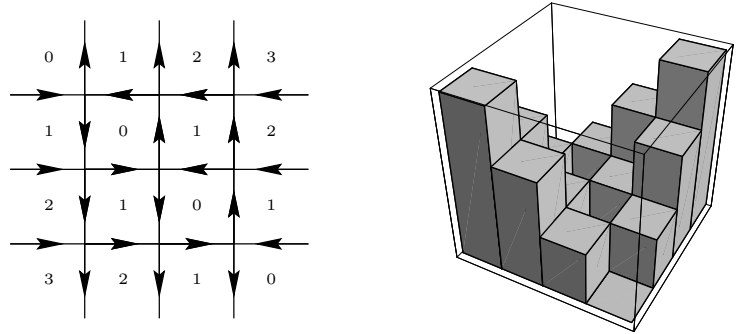}}
\end{picture}}
\caption{Vertex and height interpretation corresponding to the
$3\!\times\! 3$ identity matrix.}
\label{fig:3x3ex}
\end{figure}

\index{alternating-sign matrices}
Alternating sign matrices (ASMs) were introduced by Mills, Robbins and Rumsey
\cite{MillsRR82,MillsRR83} and are matrices with  entries in $\{-1,0,1\}$ such that the entries in each column and each row add up to $1$ and the non-zero entries alternate in sign. A well known subclass of ASMs are the permutation matrices. Let us also introduce the height interpretation of an ASM. Let $A=(a_{ij})_{i,j=1}^n$ be an ASM, then define the heights $h_{ij}$ by 
\be
h_{ij} = n-i-j+2\sum_{i' \leq i,\ j'\leq j} a_{i'j'}.
\label{heights}
\ee
This rule ensures that neighbouring heights differ by one. The correspondence between the three objects is given in Figure \ref{fig:6V}.
An example of a six vertex and its corresponding height configuration
is given in Figure \ref{fig:3x3ex} for the $3\times 3$ identity matrix.

\subsection{Structure}
\label{se:structure}
\begin{figure}[h]
\centerline{\includegraphics[width=4cm]{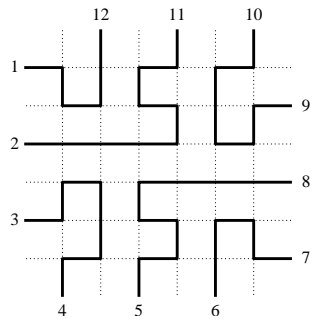}}
\caption{An FPL diagram with link pattern $((\, ()\, (())\, ()\, ))$.}
\label{fig:FPLlink}
\end{figure}

\begin{figure}[h]
\centerline{\includegraphics[width=4cm]{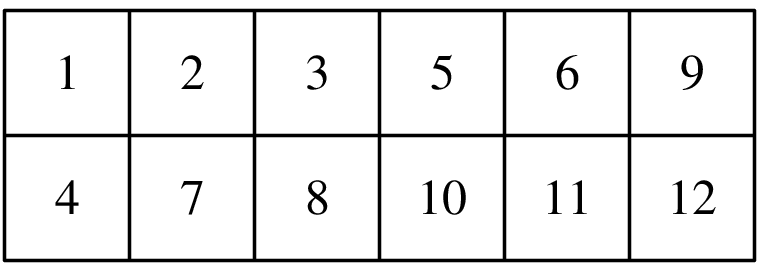}}
\caption{Standard Young tableau corresponding to the FPL diagram in
Figure~\ref{fig:FPLlink}.}
\label{fig:young}
\end{figure}

As each external terminal, or outgoing bond, is connected to another terminal, FPL diagrams can be
\index{link pattern} \index{Young tableaux} \index{Dyck paths}
naturally labeled by link patterns, or equivalently, two-row Young tableaux or Dyck paths. For example, the diagram in
Figure~\ref{fig:FPLlink} has link pattern $((\, ()\, (())\, ()\, ))$ which is short hand for saying that $1$ is connected to $12$, $2$ is connected to $11$, $3$ to $4$ etc. The information about connectivities can also be coded in two-row standard Young tableaux. The entries of the first row of the Young tableau correspond to the positions of opening parentheses '(' in a link pattern, and
the entries of the second row to the positions of the closing parentheses ')'. The FPL diagram of Figure~\ref{fig:FPLlink} carries as a label the standard Young tableau given in Figure~\ref{fig:young}.

\begin{figure}[h]
\centerline{\includegraphics[width=5cm]{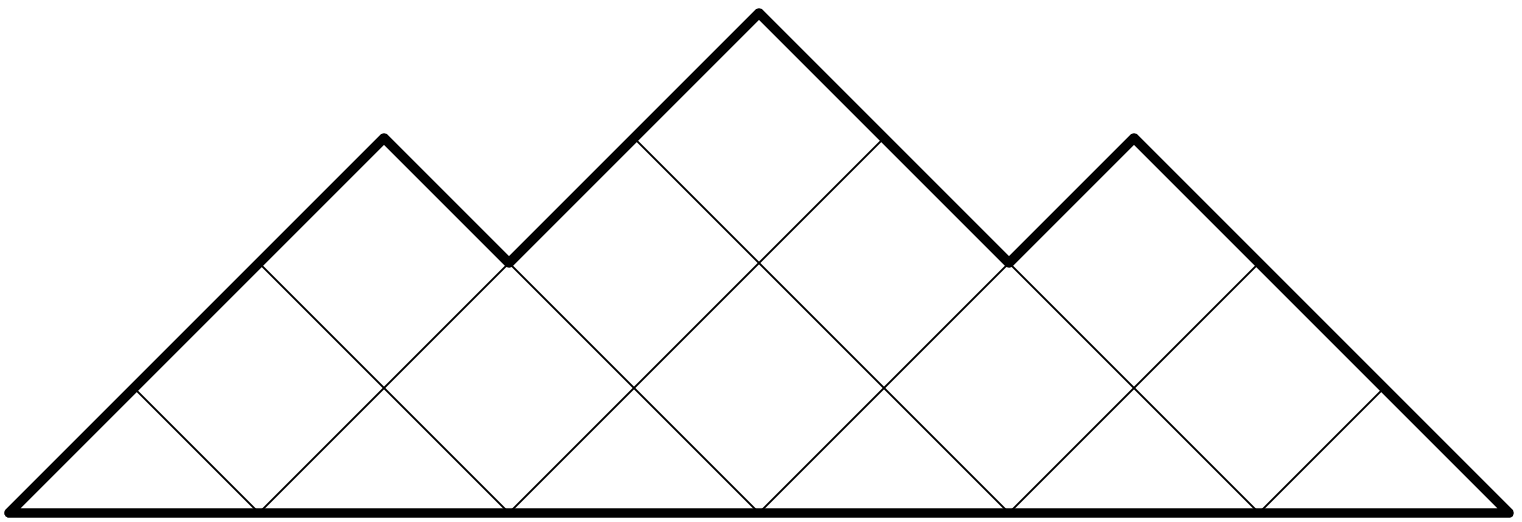}}
\caption{Dyck path corresponding to the FPL diagram in Figure~\ref{fig:FPLlink}
and the standard Young tableau in Figure~\ref{fig:young}.}
\label{fig:dyck}
\end{figure}

Yet another way of coding the same information uses Dyck paths. Each entry in the first row of the
standard Young tableau represents an up step, while those in the
second row represent down steps. The Dyck path corresponding to
Figure~\ref{fig:young} is given in Figure~\ref{fig:dyck}.

In this section we collect some structural results regarding local update moves of FPL models. 
Following Wieland \cite{Wiel00}, we define operators $G_{ij}$ that act on the height  configurations as follows. They act as the identity on each square except on the square at $(i,j)$ where they either increase or lower the height by $2$ if it is allowed. A change of height is allowed if neighbouring heights still differ by one after the change. If it is not allowed, $G_{ij}$ acts as the identity. For future convenience we also define the operators
\be
G_0 = \prod_{(i,j) \in S_0} G_{ij},\quad G_1 =
\prod_{(i,j) \in S_1} G_{ij}.
\ee
where $G_0$ and $G_1$ denote the even and odd sublattice of the square 
lattice respectively.

Starting from an initial height configuration, such as the one in
Figure \ref{fig:3x3ex}, the operators $G_{ij}$ generate all height
configurations. Put in other words, if we denote the height
configuration corresponding to the unit matrix by $Z_1$, all other
allowed height configurations correspond to a word in the operators
$G_{ij}$ acting on $Z_1$. 

On a plaquette of an FPL configuration, the involution $G$ acts as
\be
\begin{picture}(90,17)
\put(0,00){$G:$}
\put(20,-8){\includegraphics[width=20pt]{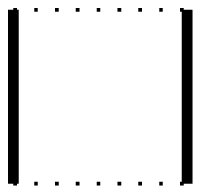}}
\put(50,0){$\leftrightarrow$}
\put(70,-8){\includegraphics[width=20pt]{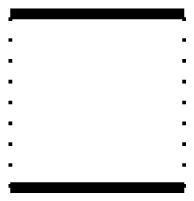}}
\end{picture}
\ee
\vskip8pt
\noindent
while on other types of plaquettes $G$ acts as the identity. Wieland \cite{Wiel00} observed that the operator $G_0 \circ G_1$ ``gyrates'' a link pattern and that the number of FPL configurations is an invariant under gyration.
 
We define two other operations on the FPL
diagrams, $U_{ij}$ and $O_{ij}$, that leave the link pattern invariant but that generate all
diagrams belonging to a fixed link pattern. The operator $U$ acts on
two plaquettes, either horizontally or vertically. Where it acts
non-trivially it is given by,
\be
\begin{picture}(130,35)
\put(0,0){$U:$}
\put(20,10){\includegraphics[width=40pt]{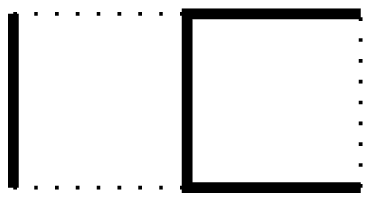}}
\put(70,18){$\leftrightarrow$}
\put(90,10){\includegraphics[width=40pt]{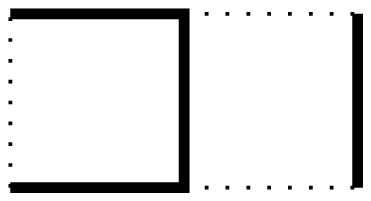}}

\put(30,-40){\includegraphics[width=20pt]{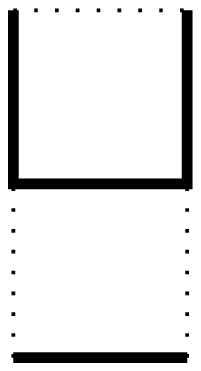}}
\put(70,-24){$\leftrightarrow$}
\put(100,-40){\includegraphics[width=20pt]{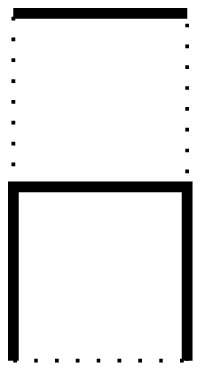}}
\end{picture}
\ee
\vskip40pt
The operator $O$ acts on three plaquettes, either horizontally or
vertically. Where it acts non-trivially, it is given by,
\be
\begin{picture}(170,45)
\put(0,0){$O:$}
\put(20,20){\includegraphics[width=60pt]{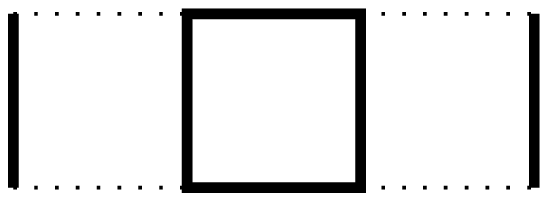}}
\put(90,28){$\leftrightarrow$}
\put(110,20){\includegraphics[width=60pt]{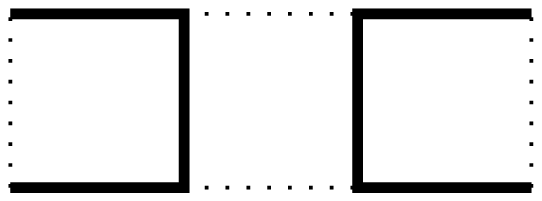}}

\put(40,-50){\includegraphics[width=20pt]{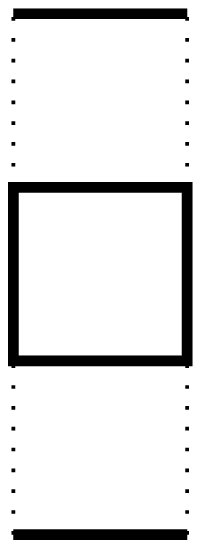}}
\put(90,-26){$\leftrightarrow$}
\put(130,-50){\includegraphics[width=20pt]{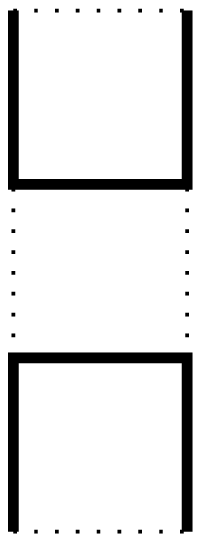}}
\end{picture}
\ee
\vskip50pt
It is easy to see that both $U$ and $O$ leave the link pattern
external to the plaquettes on which they act invariant. It is also not 
difficult to see that on a horizontal strip of arbitrary length, such
that only the leftmost and rightmost edge are connected to the outside
world, the operators $U$ and $O$ generate all possible FPL diagrams
leaving the link pattern invariant. A similar argument holds for
vertical strips. This proves that acting with $U$ and $O$ on an FPL
diagram with given link pattern, one generates all FPL diagrams
corresponding to that link pattern, and no more.

\begin{figure}[h]
\centerline{
\begin{picture}(140,70)
\put(0,0){\includegraphics[width=140pt]{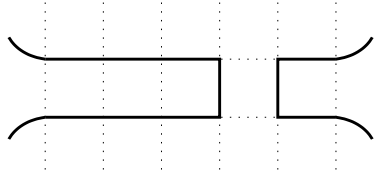}}
\end{picture}}
\caption{An isolated row inside an FPL configuration: only the leftmost
  and rightmost edge are connected to the rest of the FPL
  configuration. The operators $U$ and $O$ generate all possible
  configurations within the row.} 
\end{figure}

\section{Partition function}
\index{partition function}

\begin{figure}[h]
\centerline{
\begin{picture}(300,53)
\put(0,0){\includegraphics[width=300pt]{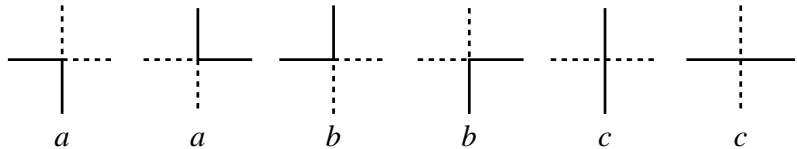}}
\end{picture}}
\caption{Weights of the six local FPL vertices.}
\label{fig:FPLweights}
\end{figure}

To each local FPL vertex we assign a weight $w_i$ and define the statistical mechanical partition function $Z_n$ as the sum over all FPL configurations of the product of the vertex weights,
\be
Z_n = \sum_{\textrm{configurations}} \prod_{i=1}^6 w_i^{k_i},
\ee
where $k_i$ is the number of vertices of type $i$. We will only consider the case where the weights on the two sublattices are the same, i.e. in the six-vertex representation the weights are invariant under arrow reversal. Using standard six-vertex notation we write $w_1=w_2=a$, $w_3=w_4=b$ and $w_5=w_6=c$, see Figure~\ref{fig:FPLweights}. 

It is convenient to parametrise $a$, $b$ and $c$ in the following way,
\be
a=\sin(\gamma-u),\qquad b=\sin(\gamma+u),\qquad c=\sin(2\gamma),
\label{weights}
\ee 
and to introduce the $u$-independent quantity $\tau$ by
\be
\tau^2 = \frac{c^2 - (a-b)^2}{ab} = 2(1-\Delta) = 4\cos^2\gamma,
\label{taudef}
\ee
where $\Delta$ is the standard notation for the anisotropy parameter of the six-vertex model defined by
\index{six-vertex model}
\be
\Delta= \frac{a^2+b^2-c^2}{2ab} = -\cos(2\gamma).
\label{Deltadef}
\ee
When $a=b$, $\tau=c/a$ gives a weight to straight loop segments. It is therefore expected that for some critical value of $\tau$ there is an ordering transition in the FPL model from a disorder phase to a phase where the vertex with weight $c$ dominates and the polygons are elongated. We will see below that this transition takes place at $\tau=2$. For $a>b+c$ or $b>a+c$ there is another ordering transition at $\tau=0$ where the vertices with weight $a$ or $b$, respectively, dominate.

The partition function $Z_n$ can be computed exactly for finite $n$ applying methods of solvable lattice models to the six-vertex model with domain wall boundary conditions. This was first done by Korepin 
\index{domain wall boundary conditions}
and Izergin \cite{Kore82,Izergin82,IzerCK92} who derived the following determinant expression for $Z_n$,
\be
Z_n = \frac{\left(\sin(\gamma+u)\sin(\gamma-u)\right)^{n^2}}{\left(\prod_{k=0}^{n-1} k!\right)^2} \ \sigma_n,
\label{Izergin}
\ee
where $\sigma_n$ is the Hankel determinant
\index{Hankel determinant}
\be
\sigma_n = \det\left( \frac{\gierd^{i+k-2} \phi}{\gierd u^{i+k-2}}\right)_{1\leq i,k \leq n},
\ee
and
\be
\phi(u) = \frac{\sin(2\gamma)}{\sin(\gamma+u)\sin (\gamma-u)}.
\ee

Using the height representation \eqref{heights} it is possible to introduce elliptic weights, rather than the trigonometric weights \eqref{weights}. The partition function in that case has been computed by Rozengren \cite{Rozen08}.

\subsection{Another form of the partition function}
Independent of Izergin and Korepin, in the case $a=b$ (i.e. $u=0$), another form of $Z_n$ was discovered conjecturally by Robbins in the context of alternating-sign matrices (ASMs) and symmetry classes thereof, see \cite{Robbins00}. As can be easily seen from Figures~\ref{fig:FPLex} and \ref{fig:6V}, a $\tau$ weighted FPL configuration, where each straight loop segment is assigned a weight $\tau$, is equal to the generating of weighted ASMs where each nonzero entry is assigned a weight $\tau$. Up to a simple factor, this is also the generating function $A_n(\tau^2)$ of $\tau^2$-weighted ASMs of size $n\times n$ where each $-1$ is assigned a weight $\tau^2$ (each additional $-1$ in an ASM also introduces an additional $+1$). The latter was conjectured by Robbins \cite{Robbins00} to equal
\be
A_{2n}(\tau^2) = 2T_{n}(\tau^2) R_{n-1}(\tau^2),\qquad A_{2n+1}(\tau^2) = T_{n}(\tau^2) R_{n}(\tau^2). 
\label{ASMRobbins}
\ee
where
\be
T_n(\tau^2) = \det_{1\leq i,j\leq n}\left( \sum_{r=0}^{2n} \gierbinom{i-1}{r-i} \gierbinom{j}{2j-r} \tau^{2(2j-r)}\right),
\label{RobbinsT}
\ee
and
\be
R_n(\tau^2) = \det_{0\leq i,j\leq n-1}\left( \sum_{r=0}^{2n-1} Y_{i,r,\mu}Y_{j,r,0}\ \tau^{2(2j+1-r)}\right),
\label{RobbinsR}
\ee
where
\be
Y_{i,r,\mu} = \gierbinom{i+\mu}{2i+1+\mu-r} + \gierbinom{i+1+\mu}{2i+1+\mu-r}.
\ee

The precise correspondence using the notation of the previous section is
\begin{subequations}
\begin{align}
Z_{2n} &= (\sin\gamma)^{2n(2n-1)}\ A_{2n}(4\cos^2\gamma),
\label{ASM2PP1}\\
Z_{2n+1} &= 2\cos\gamma\ (\sin\gamma)^{2n(2n+1)} A_{2n+1}(4\cos^2\gamma).
\label{ASM2PP2}
\end{align}
\label{ASM2PP}
\end{subequations}
In fact, Robbins' conjecture was slightly more general and gave a generating function for refined ASMs. The generating functions $R$ and $T$ appear naturally in weighted enumerations of cyclically symmetric plane partitions \cite{MillsRR87}.
\index{plane partitions!cyclically symmetric}

The equivalence of the homogeneous limit of Izergin's determinant and Robbins' conjecture, i.e. equation \eqref{ASM2PP}, is only proved for $\tau=1$ \cite{Zeilb96a,Kupe96}. Kuperberg and Robbins \cite{Kupe00,Robbins00} noticed several other such equivalences between homogeneous Izergin or Tsuchiya\footnote{The Tsuchiya determinant is the generating function of horizontally or vertically symmetric FPL diagrams \cite{Tsuch98}} type determinants and generating functions of the form (\ref{RobbinsT}) or (\ref{RobbinsR}). Some of these were recently proved in \cite{GierPZ} using a technique which seems immediately applicable to all the cases considered by Kuperberg and Robbins.

\section{Bond percolation, the O($n=1$) model and the Razumov-Stroganov conjecture}
\index{bond percolation} \index{Razumov-Stroganov conjecture}
In this section we mention a (partially conjectural) relation between FPL diagrams and the O(1) loop 
\index{O(1) loop model}
model. We will use this relation to generate FPL statistics in a relatively easy way, without having to explicitly enumerate FPL diagrams.

\begin{figure}[t]
\centerline{
\begin{picture}(200,250)
\put(0,0){\includegraphics[width=200pt]{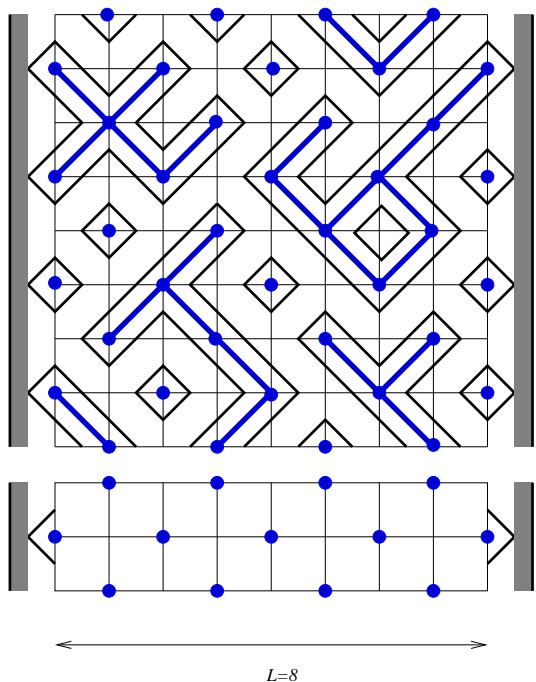}}
\end{picture}}
\caption{Bond percolation clusters and O(1) cluster boundaries on a semi-infinite strip. Configurations are generated by repeated concatenation of double rows using the double-row transfer matrix. The particular boundary conditions chosen here are called \textit{closed} or \textit{reflecting}.}
\label{fig:perc_strip}
\end{figure}

Imagine that each site of the square lattice is a reservoir of water. With probability $p$, water percolates between reservoirs along a bond of the square lattice. At $p=1/2$, the model is critical, and equivalent to the dense O($n=1$) loop model \cite{BN89} on a square lattice. The loops of the O(1) model describe the boundaries of the percolation clusters, see Figure~\ref{fig:perc_strip}. Many asymptotic properties 
\index{percolation clusters}
such as critical exponents of correlation functions can be computed for the O($n$) model using Coulomb gas techniques and conformal field theory, see Chapter 14 for an exhaustive overview. More recently, geometric properties of conformally invariant loops have been analysed using the stochastic Loewner evolution (SLE), see Chapter 15 of this book.  

\index{transfer matrix}
Configurations of the O(1) loop model can be generated using a transfer matrix, see Figure~\ref{fig:perc_strip} for the particular case of \textit{closed} or \textit{reflecting} boundary conditions. 
\index{boundary conditions!closed, reflecting}
Schematically, the local blocks of the O(1) transfer matrix are given by
\be
\raisebox{-15pt}{\includegraphics[height=30pt]{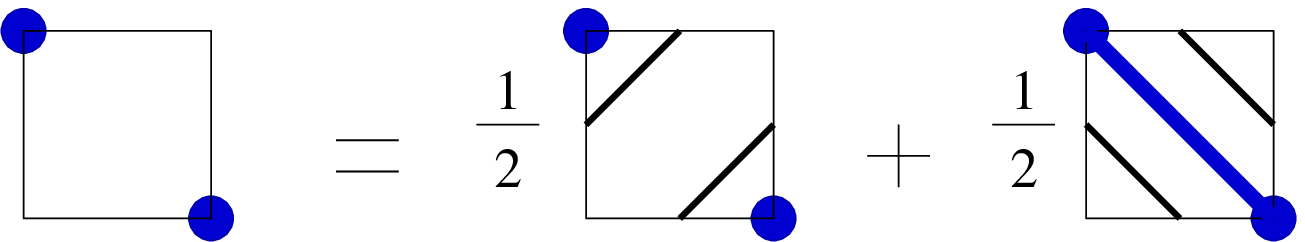}}.
\ee
The closed loops of the O(1) loop model have weight $n=1$. Loops ending on the boundary of the strip define a link pattern. For example, the link pattern corresponding to the bottom side of Figure~\ref{fig:perc_strip} has link pattern $()()()()$. The transfer matrix $T$ of the O(1) loop model therefore acts on states indexed by a link pattern. 

\subsection{The Razumov-Stroganov conjecture}
\index{Razumov-Stroganov conjecture}
\label{sec:RS}
The largest eigenvalue of the transfer matrix of the O(1) has eigenvalue 1. It was found in \cite{BatchGN01,RazuS01b,RazuS01c,GierNPR02,PearceRGN02} that the corresponding groundstate 
\index{groundstate eigenvector}
eigenvector surprisingly is related to the statistics of FPL models. Denoting a link pattern by $\alpha$ and forming a vector space with basis elements $\kett{\alpha}$, the groundstate eigenvector satisfies
\be
T\kett\psi = \kett\psi,\qquad \kett\Psi = \sum_\alpha \psi_\alpha \kett\alpha.
\label{O(1)eig} 
\ee
In the case of periodic boundary conditions, Razumov and Stroganov formulated the following important conjecture:\medskip\\
\fbox{\begin{minipage}{\textwidth}
The coefficient $\psi_\alpha$ equals the number of FPL diagrams with link pattern $\alpha$. 
\end{minipage}}
\medskip

The RS conjecture generalises to other boundary conditions, in which case the eigenvector coefficient $\psi_\alpha$ of the corresponding transfer matrix enumerates symmetry classes of FPL diagrams, to be discussed below. This is explained in detail in \cite{Gier01}. The case that will be treated in most detail here is the O(1) model on a strip, as in Figure~\ref{fig:perc_strip}, for which $\psi_\alpha$ conjecturally enumerates horizontally symmetric FPL diagrams.

Assuming the RS conjecture, we will use the O(1) loop model to generate FPL statistics by solving \eqref{O(1)eig}, and variants thereof for other boundary conditions. The particular boundary conditions we will use are \emph{periodic}, \emph{cylindrical} and \emph{closed}. See e.g. \cite{MitraNGB02,Gier01,Duchon,Pyatov,ZJ07} for other types of boundary conditions not considered here. 

Let us define the norm $\gierN_L$ of $\kett\Psi$ by
\be
\gierN_L=\sum_\alpha \psi_\alpha,
\label{norm}
\ee
and denote the largest element of $\kett\Psi$ by $\psi_{\rm max}$. The result of solving (\ref{O(1)eig}) for various boundary conditions is
\be
\renewcommand{\arraystretch}{1.6}
\begin{array}{@{}l|c|c@{}}
{\rm Type} & \gierN_L & \psi_{\rm max}\\ \hline\hline
{\rm Periodic},\; L\; {\rm even} & A(L/2) & A(L/2-1) \\
{\rm Cylindrical},\; L\; {\rm even} & A_{\rm HT}(L) & A_{\rm HT}(L-1)\\
{\rm Cylindrical},\; L\; {\rm odd} & A_{\rm HT}(L)& A((L-1)/2)^2 \\
{\rm Closed},\; L\; {\rm even} & A_{\rm V}(L+1) & A_{\rm V}(L) \\
{\rm Closed},\; L\; {\rm odd} & A_{\rm V}(L+1) & A_{\rm V}(L)
\end{array}
\label{sumrules}
\ee
where the numbers $A$, $A_{\rm HT}$ and $A_{\rm V}$ are defined by,
\begin{itemize}
\item
The number of $n\times n$ ASMs,
\begin{equation}
A(n) = \prod_{k=0}^{n-1} \frac{(3k+1)!}{(n+k)!} \;=\;
1,2,7,42,\ldots
\end{equation}
\item
The number of $n\times n$ half turn symmetric ASMs,
\index{alternating-sign matrices!half turn symmetric}
\begin{equation}
\renewcommand{\arraystretch}{1.6}
\begin{array}{l}
\displaystyle A_{\rm HT}(2n) = A(n)^2\prod_{k=0}^{n-1} \frac{3k+2}{3k+1} \; =\; 2,10,140,5544,\ldots\\
\displaystyle A_{\rm HT}(2n-1) = \prod_{k=1}^{n-1} \frac{4}{3}
\left(\frac{(3k)!(k!)}{(2k)!^2}\right)^2 \; =\; 1,3,25,588,\ldots
\end{array}
\end{equation}
\item
The number of $(2n-1) \times (2n-1)$ horizontally (or vertically) symmetric ASMs, 
\index{alternating-sign matrices!horizontally, vertically symmetric}
\begin{equation}
A_{\rm V}(2n-1) = \prod_{k=1}^{n-1} (3k-1)
\frac{(6k-3)!(2k-1)!}{(4k-2)!(4k-1)!} \; =\; 1,1,3,26,646,\ldots.
\end{equation}
and its related version for even sizes (also denoted by $N_8$ in \cite{Bress99}),
\begin{equation}
A_{\rm V}(2n) = \prod_{k=1}^{n-1} (3k+1)
\frac{(6k)!(2k)!}{(4k)!(4k+1)!} \; =\; 1,2,11,170,\dots
\end{equation}
\end{itemize}

Mills et al. conjectured the number of ASMs to be $A(n)$, which was proved more than a decade later by Zeilberger \cite{Zeilb96a} and in an entirely different way by Kuperberg \cite{Kupe96}. Kuperberg made essential use of the connection to the six-vertex model and its integrability. Conjectured enumerations of 
\index{six-vertex model}
symmetry classes were given by Robbins \cite{Robbins00}, many of which were subsequently proved by Kuperberg \cite{Kupe00}. The properties and history of ASMs are reviewed in the book by Bressoud \cite{Bress99}, as well as by Robbins \cite{Robbins91} and Propp \cite{Propp01}.

\subsection{Proofs and other developments}
The sum rules listed in Table~\ref{sumrules}, relating the norms \eqref{norm} of $\kett\Psi$ for different boundary conditions to symmetry classes of alternating-sign matrices, were originally obtained conjecturally. These sum rules have been proved algebraically using an inhomogeneous extension of the transfer matrix, a method initiated and developed by Di Francesco and Zinn-Justin \cite{DFZJ04,DF05}. This has led to further interesting directions, not pursued here, such as the connections between weighted FPL diagrams (or ASMs), plane partitions and the $q$-deformed 
Knizhnik-Zamolodchikov equation \cite{Pasquier,DF05,DFZJ05,DF07,DFZJ07,GierPZ}.
\index{plane partitions} \index{q-deformed KZ equation}

In an alternative interpretation, the O(1) model is equivalent to a stochastic model defined on link patterns, the so called raise and peel model \cite{GierNPR04}. It is an open question how to define a 
\index{raise and peel model} \index{fully packed loops!diagrams}
stochastic model directly on FPL diagrams, by say the Wieland involutions $G$ describe in Section~\ref{se:structure}, such that it has an equipartite stationary state and reduces to the raise and peel model when the action of the operators $O$ and $U$ of Section~\ref{se:structure} is divided out. Such a process would result in a direct proof of the Razumov-Stroganov conjecture.

\section{Symmetry classes of FPL diagrams}
\index{fully packed loops!symmetry classes}

We will now focus on FPL models defined on rectangular grids, corresponding to certain symmetry classes of square FPL diagrams. The two main reasons are that for such FPL models there is a natural boundary giving rise to additional structure, and that at the time of writing, for these models more results are known which are relevant to polygon models. 

%
%

\subsection{Horizontally symmetric FPL diagrams}
\index{fully packed loops!horizontally symmetric}

For horizontally symmetric FPL diagrams (HSFPLs) one only has to consider the lower half of an FPL diagram. As explained in \cite{Gier01}, due to geometric constraints one can further reduce the size of such half diagrams. Therefore, for $L$ even, the reduced lower half of a horizontally symmetric FPL diagram of size $(L+1)\times (L+1)$ is an FPL diagram of size $(L-1)\times L/2$. The total number $Z_{\rm HSFPL}(2n)$ of horizontally (or vertically) symmetric FPL diagrams of size $(2n-1)\times n$ is known, and can be computed from the Tsuchiya determinant \cite{Tsuch98,Kupe00}, 
\index{Tsuchiya determinant}
\be
Z_{\rm HSFPL}(2n) = A_{\rm V}(2n+1) = \prod_{k=1}^{n} (3k-1)
\frac{(6k-3)!(2k-1)!}{(4k-2)!(4k-1)!}.
\ee

\begin{figure}[h]
\centerline{
\begin{picture}(140,110)
\put(0,0){\includegraphics[width=140pt]{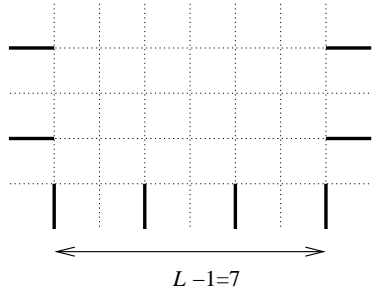}}
\end{picture}}
\caption{Boundary conditions for an HSFPL diagram of size $(2n-1)\times n=7\times 4$. The number of external terminals equals $2n=8$, hence the statistics of this diagram is generated from the O(1) model with $L=8$.}
\label{fig:HSFPL8}
\end{figure}

As can be seen from Table~\ref{sumrules}, this number is equal to the norm $\gierN_{2n}$ for the O(1) model with closed boundary conditions and $L=2n$. For odd system sizes, $L=2n+1$, the norm $\gierN_{L}$ equals the number of FPL diagrams of size $L\times (L-1)/2$, which we will denote by $Z_{\rm HSFPL}(2n+1)$.

\begin{figure}[t]
\centerline{
\begin{picture}(250,170)
\put(0,0){\includegraphics[width=250pt]{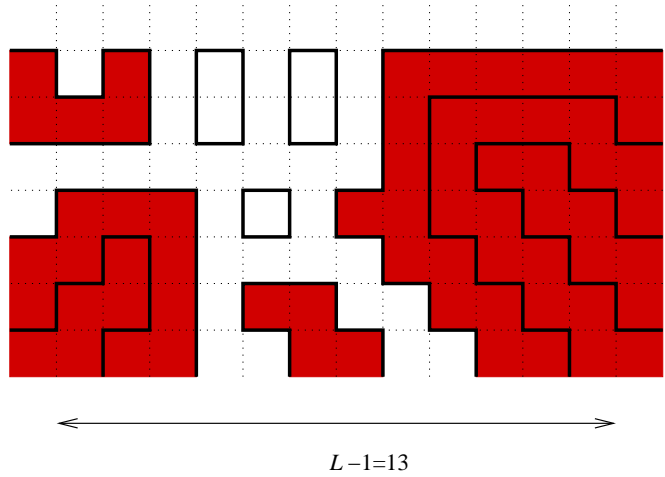}}
\end{picture}}
\caption{An FPL diagram of size $(L-1)\times L/2=13\times 7$ with four nests.}
\label{fig:nestsV}
\end{figure}

There are two interesting and natural statistics on HSFPLs which we will explain now. As noted above, 
\index{link pattern}
to each FPL diagram is associated a link pattern. Each link pattern factorises in sets of completed links where, in terms of the parenthesis notation, the number of closing parentheses equals the number of opening parentheses. For example,
\[
(()())(((())())() = (()())\cdot (((())()) \cdot ().
\] 
Such completed links are called nests \index{nest}, and they provide a statistic for HSFPLs. An example of an HSFPL diagram of size $13\times 7$ with four nests is given in Figure~\ref{fig:nestsV}.

\begin{figure}[h]
\centerline{
\begin{picture}(280,160)
\put(0,0){\includegraphics[width=270pt]{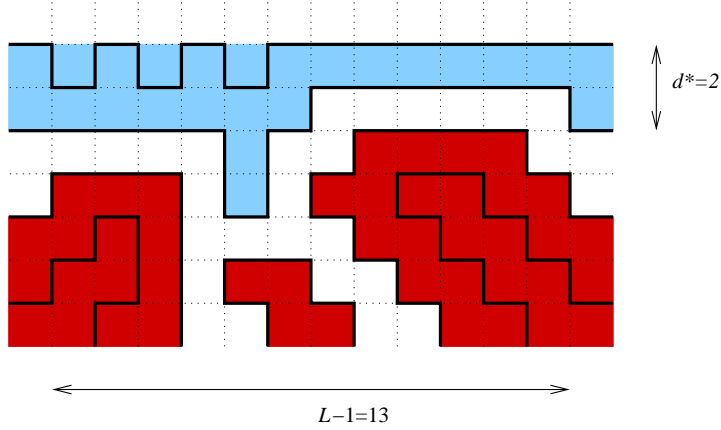}}
\end{picture}}
\caption{An FPL diagram of size $(L-1)\times L/2=13\times 7$ with three nests and depth $d=4$ ($d^*=2$).}
\label{fig:nestsV2}
\end{figure}

Another natural statistic is the number $d^*$ of loops connecting the leftmost loop terminals with the rightmost ones, i.e. loops connecting terminal $i$ with $2\lfloor L/2\rfloor -i+1$ for $i=1,\ldots,d^*$. It will be convenient to define $d$ by
\be
d =\left\lfloor \frac{L-1}{2}\right\rfloor -d^*,
\ee
where $d$ is called the depth of an HSFPL diagram. An example of an HSFPL diagram of size $(L-1)\times L/2=13\times 7$ with three nests and depth $d=4$ ($d^*=2$) is given in Figure~\ref{fig:nestsV2}. 

\subsection{Depth-nest enumeration of HSFPLs} 
\index{depth-nest enumeration}
In this section we will say that an FPL diagram is of size $L$, if it is of size $(L-1)\times L/2$ if $L$ is even, or of size $L\times (L-1)/2$ if $L$ is odd. Let $P(L,d,m)$ be the number of such FPL diagrams of size $L$, depth $d$ and having $m+1$ nests. The nest generating function for diagrams of size $L$ and 
\index{nest!generating function}
depth $d$ is defined by
\be
\gierP(L,d;z) = \sum_{m=0}^{d} P(L,d,m) z^m.
\ee 
Let $S(L,d)$ be the total number of HSFPL diagrams at a given size $L$ and depth $d$. Obviously we have 
\be
S(L,d) = \gierP(L,d;1) = P(L,d+1,0),
\ee
and
\be
Z_{\rm HSFPL}(L) = S(L,\lfloor \tfrac{L-1}{2}\rfloor)=\gierP(L,\lfloor \tfrac{L-1}{2}\rfloor;1).
\ee
Based on the RS conjecture, Mitra et al. and Pyatov have conjectured the exact form of $S(L,d)$ \cite{MitraNGB02,Pyatov}. Here we give this conjecture in the following form:
\begin{conjec}
\label{conj:NestTot}
The total number of HSFPL diagrams at a given size $L$ and depth $d$ is given by
\be
S(L,d) = \prod_{k=0}^{d} \frac{\Gamma(L-k+1)}{2^k(1/2)_k \Gamma(L-2k+1)}\frac{ \Gamma(\frac{2L+2k+3}{6}) \Gamma(\frac{L-2k+3}{3})}{\Gamma(\frac{2L-k+3}{6}) \Gamma(\frac{2L-k+6}{6})}.
\label{SLd}
\ee
\end{conjec}
Assuming the RS conjecture, the formula for $S(L,d)$ has recently been proved \cite{GierPZ}.

Pyatov also found an exact formula for $P(L,d,m)$ \cite{Pyatov} which fits exact data for small system sizes ($L\leq 18$). He conjectured that this formula holds for all $L$, $d$ and $m$. In terms of the nest generating function this conjecture can be stated as follows.
\begin{conjec}
\label{conj:NestGen}
The nest generating function is given by
\index{nest!generating function}
\be
\gierP(L,d;z) = S(L,d-1) \ _3F_2\!\left(
\begin{array}{c}-d,L-2d,L-d+\frac{1}{2} \\
-2d,2 L-2 d+1
\end{array};4
   z\right).
\ee
\end{conjec}
Note that Conjecture~\ref{conj:NestTot} follows from Conjecture~\ref{conj:NestGen} due to the evaluation
\be
\ _3F_2\!\left(
\begin{array}{c}-d,L-2d,L-d+\frac{1}{2} \\
-2d,2 L-2 d+1
\end{array};4\right) = \frac{S(L,d)}{S(L,d-1)},
\ee
which is a consequence of one of the strange evaluations \index{strange evaluations}
of Gessel and Stanton \cite{GesselS}. For $d=\lfloor (L-1)/2\rfloor$, the formulas in Conjecture~\ref{conj:NestTot} and Conjecture~\ref{conj:NestGen} were given in \cite{Gier01}.

By convention, $P(L,d,m)$ have the following boundary values:
\be
\label{split-2}
P(L,d,m=-1) = P(L,d,m=-2) = P(L,d,m=d+1) =0\, ,
\ee
and we also note the boundary condition
\be
\label{split-4}
P(L,d,m=1) = (L-2d) S(L,d-1)\, .
\ee
It was found in \cite{APR} that the function $P(L,d,m)$ is completely determined by these boundary conditions and the following interesting bilinear relation called the \textit{split hexagon relation},
\index{split hexagon relation}
\be
\renewcommand{\arraystretch}{1.6}
\begin{array}{l}
P(L+1,d+1,m) S(L-1,d-1) \\
\hphantom{P(L }= P(L-1,d,m) S(L+1,d) + P(L,d-1,m-2) S(L,d+1).
\end{array}
\label{split-3}
\ee
Summing up over $m=0,1,\dots d+1$ in (\ref{split-3}) reproduces the hexagon relation, or discrete Boussinesq equation, for $S(L,d)$, see \cite{Pyatov}.
\index{discrete Boussinesq equation}

\subsubsection{Cyclically symmetric transpose complement plane partitions}
\begin{figure}[h]
\centerline{
\begin{picture}(250,250)
\put(0,0){\includegraphics[width=250pt]{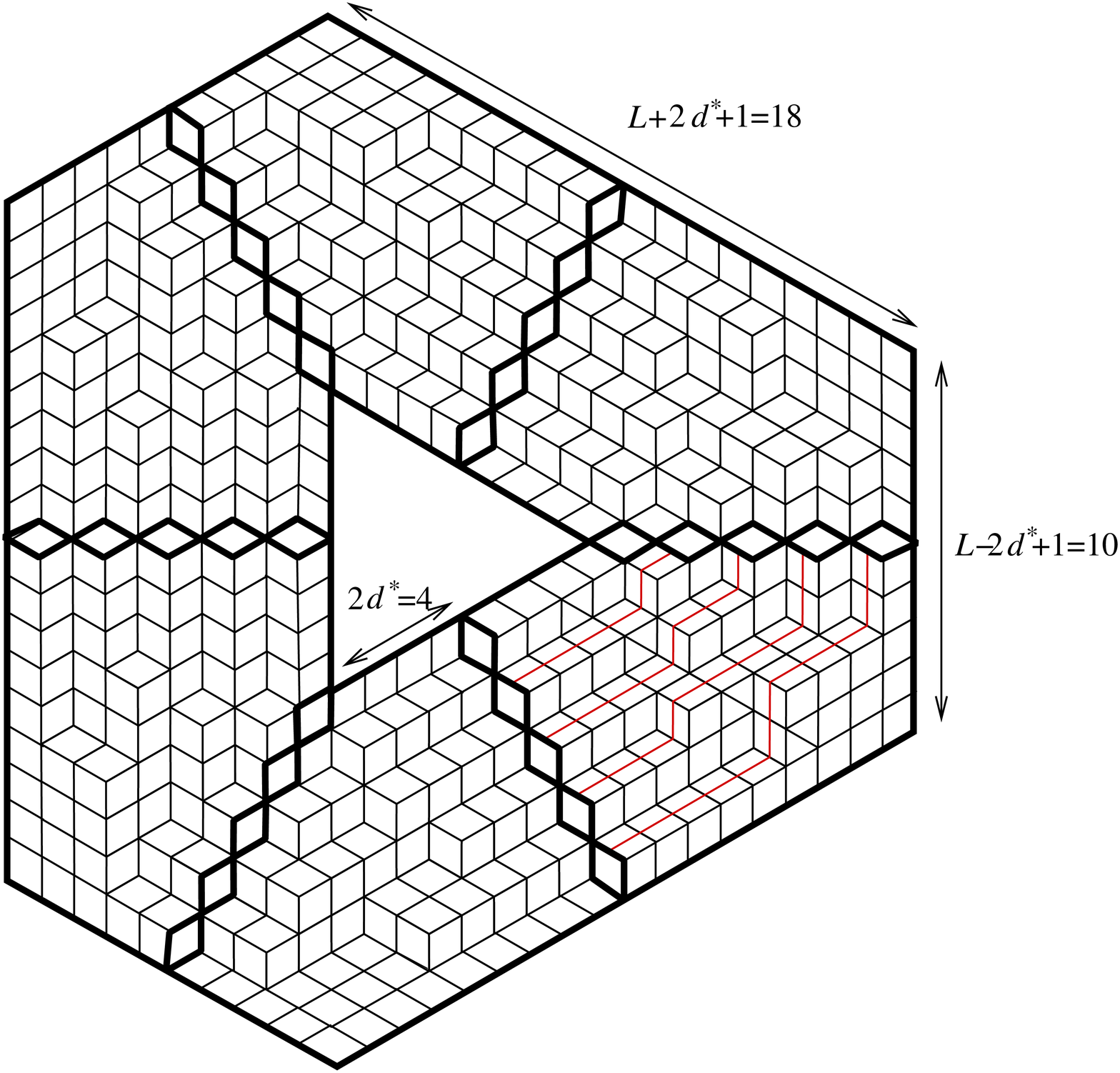}}
\end{picture}}
\caption{A punctured cyclically symmetric transpose complement plane partition for $L=13$ and $d^*=2$.}
\label{fig:PCSTCPP}
\end{figure}

\index{plane partitions!cyclically symmetric}
Somewhat outside the scope of this book, we note the following interesting fact observed in \cite{GierPZ}. The total number of nests at a given depth, $S(L,d)$, is equal to the number of punctured cyclically symmetric transpose complement plane partitions \cite{CK}, see Figure~\ref{fig:PCSTCPP}. This can be seen by enumerating the number of non-intersecting lattice paths in the South-East fundamental domain of the plane partition. 
\index{plane partitions!punctured}
Using the Gessel-Viennot-Lindstr\"om method 
\index{Gessel-Viennot-Lindstr\"om method}
\cite{GesselV,Lindstrom} one obtains a determinant of the type \eqref{RobbinsT} with $\tau=1$, which can be evaluated in factorised form \cite{CK}. This form equals the expression in \eqref{SLd}. 

\subsection{Average number of nests in HSFPL diagrams}
\label{sec:avnests}
The average number of nests in HSFPL diagrams at depth $d$ and size $L$, denoted by, $\langle 1+m\rangle_{d^*}$, is defined as
\begin{align}
 \langle 1+m\rangle_{d^*} = \frac{1}{Z_{\rm HSFPL}(L)} \sum_{m=0}^d (1+m) P(L,d,m) \nonumber\\ \equiv \frac{S(L,d)}{Z_{\rm HSFPL}(L)}\left(1+\langle m\rangle_{d^*}^{\rm c}\right).
\end{align}
For notational clarity we will suppress the dependence of $\langle 1+m\rangle_{d^*}$ on $L$ and recall that 
\[
d=\left\lfloor \frac{L-1}{2} \right\rfloor -d^*.
\]
With the data $P(L,d,m)$ we can calculate $\langle m\rangle_{d^*}^{\rm c}$:
\begin{align}
\langle m\rangle_{d^*}^{\rm c}
&= \frac{1}{S(L,d)}\sum_{m=1}^{d} m \, P(L,d,m)\nonumber\\
&= \left.\frac{\gierd}{\gierd z}\right|_{z=1} \log \gierP(L,d;z).
\label{split-10}
\end{align}

The \texttt{Mathematica} implementation of the Gosper-Zeilberger algorithm \
\index{Gosper-Zeilberger algorithm}
\cite{Gosper,Zeilb90,Zeilb91} by Paule and Schorn \cite{PauleS}, is able to recognise $\langle m\rangle_{d^*}$ in an almost factorised form. Let $L=2n$, then define $\mu_n(d^*)$ by 
\be
\renewcommand{\arraystretch}{2.2}
\begin{array}{l}
\displaystyle \mu_n(d^*) = \sum_{m=0}^{n-1-d^*} \Bigl(3m+4(d^*+1)\Bigr) \frac{P(2n,n-1-d^*,m)}{P(2n,n-1-d^*,0)} \\
\displaystyle \hphantom{\mu_n(p)} = \Bigl(3 \langle m\rangle_{d^*}^{\rm c} +4(d^*+1)\Bigr) \frac{S(2n,n-1-d^*)}{S(2n,n-2-d^*)}.
\end{array}
\ee
The expression $\mu_n(d^*)$ turns out to be summable in factorised form, giving rise to
\be
\langle m\rangle_{d^*}^{\rm c} = -\frac23(L-2d) + 2^{2/3} \frac{\Gamma\left(\frac{2L+2d+5}{6}\right)\Gamma\left(\frac{2L-d+3}{3}\right)\Gamma\left(\frac{L-2d+1}{3}\right)}{\Gamma\left(\frac{2L+2d+3}{6}\right)\Gamma\left(\frac{2L-d+2}{3}\right)\Gamma\left(\frac{L-2d}{3}\right)}.
\label{avm}
\ee
This formula also holds for odd values of $L$.

\subsection{Half-turn symmetric FPL diagrams}
\begin{figure}[h]
\centerline{
\begin{picture}(180,160)
\put(0,0){\includegraphics[width=180pt]{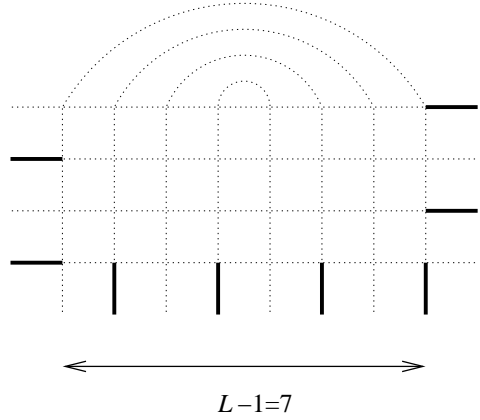}}
\end{picture}}
\caption{Boundary conditions for an HTSFPL diagram of size $(2n-1)\times n=7\times 4$. The arcs at the top are additional edges which may contain loop segments. The number of external terminals equals $2n=8$, hence the statistics of this diagram is generated from the periodic O(1) model with $L=8$.}
\label{fig:HTSFPL8}
\end{figure}

\index{fully packed loops!half-turn symmetric}
In the case of half turn symmetric FPL diagrams (HTSFPLs) it also suffices to consider only the lower half of an FPL diagram, but the boundary conditions on the top row of the half diagram are different from HSFPLs, see Figure~\ref{fig:HTSFPL8}. The total number of HTSFPL diagrams is given by \cite{Kupe00}
\begin{equation}
Z_{\rm HT}(2n) = A_{\rm HT}(2n) = 2 \prod_{k=1}^{n-1} \frac{3
(3k+2)!(3k-1)!k!(k-1)!} {4(2k+1)!^2(2k-1)!^2}.
\end{equation}

\begin{figure}[h]
\centerline{
\begin{picture}(180,160)
\put(0,0){\includegraphics[width=180pt]{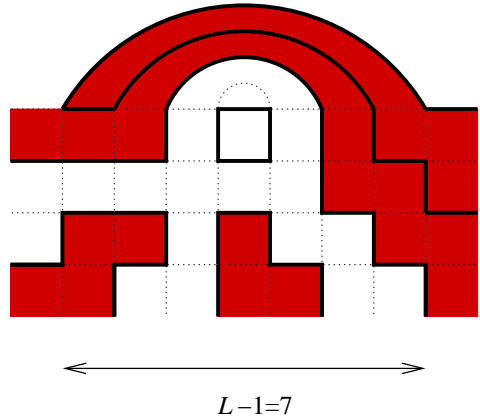}}
\end{picture}}
\caption{A HTSFPL diagram of size $(2n-1)\times n=7\times 4$ with link pattern $)\cdot ()\cdot ()\cdot (()$, having three nests.}
\label{fig:HTSFPL8_example}
\end{figure}

Care has to be taken when defining link patterns and nests for HTSFPL diagrams. External terminals can be connected in two distinct ways depending on whether the corresponding loop runs over an odd or even number of the arcs on the top of the diagram. In the case of an odd number of arcs, we exchange the parentheses denoting the connection of a pair of sites. For example, the connectivity of the HTSFPL diagram in Figure~\ref{fig:HTSFPL8_example} is denoted by 
\[ )\cdot ()\cdot ()\cdot ((),
\]
where the dots again denote the factorisation of link pattern into nests. Figure~\ref{fig:HTSFPL8_example} thus denotes an HTSFPL diagram with three nests.

As in the case of horizontal symmetry, there exists a conjecture for the nest distribution function \cite{Gier01}, but in this case only for $L=2n$ and $d^*=0$. Let $P(L,m)$ denote the number of half-turn symmetric FPL diagrams with $m+1$ nests, and define the nest generating function by
\be
\gierP(L;z) = \sum_{m=0}^{n-1} P(L,m) z^m.
\ee
%
%
\begin{conjec}
\label{con:HTnests}
\index{nest!generating function}
The nest generating function for half-turn symmetric FPL diagrams is given by
\[
\gierP(2n;z) = Z_{\rm HTSFPL}(2n) \frac{3 n}{4n^2-1}\ _3F_2\left( 
\begin{array}{@{}c}
3/2,1-n,1+n \\
2-2n,2+2n
\end{array};4z\right).
\]
\end{conjec}

The average number of nests in HTSFPL diagrams of size $L=2n$ having $1+m$ nests, denoted by $\langle 1+m\rangle$, is defined as
\begin{align}
 \langle 1+m\rangle = \frac{1}{Z_{\rm HTSFPL}(L)} \sum_{m=0}^{n-1} (1+m) P(L,d,m) = 1+ z \frac{\gierd}{\gierd z} \log \gierP(L;z)
\end{align}
Knowing the nest generating function we may compute $\langle 1+m\rangle$, which turns out to be summable \cite{Gier01}.
\begin{conjec}
\label{con:avnestHT}
The average number of nests in  HTSFPL diagrams of size $L$, is given by
\[
\langle 1+m\rangle = n \prod_{j=1}^{n-1} \frac{3j+1}{3j+2}.
\]
\end{conjec}

\section{Phase transitions}
\index{phase transition}

\subsection{Bulk asymptotics and phase diagram}

\index{fully packed loops!phase diagram}
The phase diagram of the FPL model can be derived from the asymptotics of the partition function $Z_n$ defined in (\ref{Izergin}). The leading asymptotics of $Z_n$ for general values of $\tau$ has been computed by Korepin and Zinn-Justin \cite{KoreZ} using the Toda equation satisfied by $\sigma_n$ 
\index{Toda equation}
\cite{Szogo},
\be
\sigma_n \frac{\gierd^2\sigma_n}{\gierd u^2} - \left(\frac{\gierd\sigma_n}{\gierd u}\right)^2 = \sigma_{n+1}\sigma_{n-1}.
\ee
Writing $\sigma_n$ as a matrix model integral \cite{Zinn}, further subleading asymptotics were computed 
\index{matrix model} \index{asymptotics}
by Bleher and Fokin \cite{BleherF} and Bleher and Liechty \cite{BleherL} using orthogonal polynomials. For special values of $\gamma$ this method was first employed by Colomo and Pronko \cite{ColoP}. The final result for $0< \tau^2=4\cos^2\gamma < 4$\ ($1>\Delta >-1$) is that for some $\varepsilon>0$,
\be
Z_n = C n^{\kappa} \exp\left[f n^2 (1+\mathcal{O}(n^{-\varepsilon}))\right],
\label{ZDisorder}
\ee
where $C$ is a constant and
\begin{align}
f &= \frac{\pi\ \sin(\gamma+u)\sin(\gamma-u)}{2\gamma\cos(\pi u/2\gamma)},\\
\kappa &= \frac{1}{12} - \frac{2\gamma^2}{3\pi(\pi-2\gamma)}.
\end{align}

\begin{figure}[h]
\centerline{
\begin{picture}(140,140)
\put(0,0){\includegraphics[width=140pt]{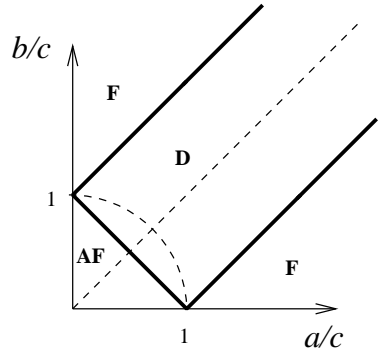}}
\end{picture}}
\caption{Bulk phase diagram of the FPL model. The phases are traditionally called disordered (D), ferro-electric (F) and anti-ferro-electric (AF), cf. the six-vertex model. The arc corresponds to the free-fermion condition $\Delta=0$ ($\tau^2=2$), and the line $a=b$ corresponds to $\tau^2$-enumerations of ASMs. On this line $\tau^2=c^2/a^2=2(1-\Delta)$, and the D-AF phase transition takes place at $\tau^2=4$.} 
\label{fig:6Vphase}
\end{figure}

The result (\ref{ZDisorder}) is valid in the so-called disordered (D) phase $0< \tau^2< 4$. There is a phase transition to an ordered phase at $\tau^2=4$ where the vertices with weight $c$ are favoured and the perimeters of the polygons in the FPL model consist of elongated straight lines. In terms of the six-vertex model this is called the anti-ferromagnetic (AF) phase. At $\tau=0$, i.e. $a=b+c$ or $b=a+c$, 
\index{phase!anti-ferromagnetic} \index{phase!ferromagnetic}
there is another phase transition  to a so-called ferromagnetic phase, where, respectively, the $a$- or $b$-type vertices dominate. The complete phase diagram is given in Figure~\ref{fig:6Vphase}

\index{phase transition!bulk}
The phase diagram and the Bethe-Ansatz solution of the six-vertex model for \textit{periodic} and
\index{Bethe Ansatz solution} \index{six-vertex model}
\textit{anti-periodic} boundary conditions are thoroughly discussed in the works of Lieb \cite{Lieb1,Lieb2,Lieb3}, Lieb and Wu \cite{LiebWu}, Sutherland \cite{Suth}, Baxter \cite{Baxter82}, and Batchelor et al. \cite{BatchBRY}.

\subsection{Asymptotics for symmetry classes at $\tau=1$}
In this section we determine the asymptotics of equally weighted horizontally and half-turn symmetric FPL diagrams for $\tau=1$ or $\gamma=2\pi/3$, corresponding to the numbers given in Table~\ref{sumrules}. The leading asymptotic form of these numbers, which are all products over factorials, can be computed using the Euler-Maclaurin approximation. Full asymptotics can easily be 
\index{Euler-Maclaurin approximation} \index{Barnes' $G$-function}
derived using Barnes' $G$-function \cite{Barnes}, which satisfies
\be
G(z+1)=\Gamma(z)G(z),\qquad G(1)=1,
\ee
and whose leading asymptotic behaviour is given by (see e.g. \cite{Mitra07}),
\be
\log(G(z+1)) = z^2\left(\frac12 \log z -\frac34\right) + \frac12 z \log 2\pi - \frac{1}{12} \log z +\mathcal{O}(1). 
\ee
In the case of $A(n)$ and $A_{\rm HT}(n)$, a detailed asymptotic analysis including the lower order terms was carried out by Mitra and Nienhuis \cite{MitraN04}. Here we list only the leading asymptotics of the FPL numbers relevant to the current context. The generic asymptotic form of the numbers is 
\begin{equation}
\log Z_L = s_0\; {\rm Area} + f_0\; {\rm Surface} + x \log ({\rm Length}) + \mathcal{O}(1),
\end{equation}
where the bulk and boundary entropies are given by
\begin{equation}
s_0 = \log \left(\frac{3\sqrt{3}}{4}\right),\qquad
f_0 = \log \left(\frac{3\sqrt{3}}{4\sqrt{2}}\right).
\end{equation}
The critical exponent $x$ is a universal quantity. In detail, the cases relevant for this chapter are
\begin{itemize}
\item{FPL diagrams, $L$ even}

The number $A(L/2)$ counts FPL configurations on an $L/2\times L/2$ square grid, the area of
which is $\frac{1}{4} L^2$. We thus find,
\begin{equation}
\log Z(L) = \log A(L/2) = \frac{1}{4} s_0 L^2 - \frac{5}{36} \log L +
O(1). 
\end{equation}
\item{Half turn symmetric FPL diagrams, $L$ even}

$A_{\rm HT}(L)$ counts the number of FPL configurations on half an
$L\times L$ square grid, the area of which is $\frac{1}{2} L^2$. We
thus find for $L$ even,  
\begin{equation}
\log Z_{\rm HT}(L) = \log A_{\rm HT}(L) = \frac{1}{2} s_0 L^2 + \frac{1}{18}
\log L + O(1).
\end{equation}
\item{Half turn symmetric FPL diagrams, $L$ odd}

$A_{\rm HT}(L)$ counts the number of FPL configurations on a square grid of dimension
$L\times (L-1)/2$, the area of which is $\frac{1}{2} L(L-1)$. We
thus find for $L$ odd,  
\begin{equation}
\log Z_{\rm HT}(L) = \log A_{\rm HT}(L) = \frac{1}{2} s_0 L(L-1) + \frac{1}{36}
\log L^2 + O(1).
\end{equation}
\item{Horizontally symmetric FPL diagrams, $L$ even}

$A_{\rm V}(L+1)$ counts the number of FPL configurations on an $(L-1)
\times L/2 $ rectangular grid. We find,       
\begin{equation}
\log Z_{\rm HSFPL}(L) = \log A_{\rm V}(L+1) = \frac{1}{2} s_0 L(L-1) + f_0 L -
\frac{5}{72} \log L + O(1).
\end{equation}
\item{Horizontally symmetric FPL diagrams, $L$ odd}

For $L$ odd, $A_{\rm V}(L+1)$ counts the number of FPL configurations
on a $L \times (L-1)/2 $ rectangular grid. We find,      
\begin{equation}
\log Z_{\rm HSFPL}(L) = \log A_{\rm V}(L+1) = \frac{1}{2} s_0 L(L-1) + f_0 L +
\frac{7}{72} \log L + O(1).
\end{equation}
\end{itemize}
Note that because the upper boundary for FPL diagrams corresponding to HSFPLs is not fixed, see e.g. Figure~\ref{fig:HSFPL8}, there is a nonzero boundary entropy in $\log Z_{\rm HSFPL}(L)$.

\subsection{Nest phase transitions}
\index{phase transition!nest}
\index{nest!phase transition}
From Section~\ref{sec:avnests} we recall that the average number of nests is given by $\frac{S(L,d)}{Z_{\rm HSFPL}(L)}(1+\langle m\rangle_{d^*}^{\rm c})$ where
\begin{equation}
\langle m\rangle_{d^*}^{\rm c} = z \frac{\D}{\D z} \log \gierP(L,d;z), \label{eq:dens}
\end{equation}
with $\gierP(L,d;z)$ given in Conjecture~\ref{conj:NestGen}. The asymptotics for $\langle m\rangle_{d^*}^{\rm c}$ as $L\rightarrow \infty$ can be derived from the hypergeometric equation satisfied by $\gierP(L,d;z)$. Taking $L=2n$ this gives
\begin{multline}
\theta(\theta+1+2d^*-2n)(\theta+2+2d^*+2n)\gierP(2n,d;z) = \\
4z(\theta+2+d^*)(\theta+1+d^*-n)(\theta+3/2+n) \gierP(2n,d;z), \label{eq:ohyp}
\end{multline}
where $\theta = z\, \D/\D z$ and
\be
d=\left\lfloor\frac{L-1}{2}\right\rfloor -d^* = n-1-d^*.
\label{d2d*}
\ee
Assuming that $d^*=\mathcal{O}(1)$, we discriminate the cases $z<1$, $z=1$ and $z>1$.
\begin{itemize}
\item{$z<1$}

In this case, up to an overall constant factor, the leading asymptotics of $\gierP(2n,d;z)$ will be polynomial in $n$. Neglecting lower order terms, equation (\ref{eq:ohyp}) reduces to,
\begin{equation}
\theta \gierP(2n,d;z) = z(\theta +2+d^*)\gierP(2n,d;z).
\end{equation}
We thus we find $(1-z)\gierP'(2n,d;z) = (2+d^*)\gierP(2n,d;z)$ and
\begin{equation}
\langle m\rangle_{d^*}^{\rm c} = (2+d^*)\frac{z}{1-z}\qquad (n\rightarrow \infty).
\end{equation}
\item{$z=1$}

In \eqref{avm} an exact expression was given for $\langle m\rangle_{d^*}^{\rm c}$ at $z=1$. Asymptotically we find that for $L-2d={\mathcal O}(1)$,
\be
\langle m \rangle_{d^*}^{\rm c} \approx \frac{\Gamma(\frac{L-2d+1}{3})}{\Gamma(\frac{L-2d}{3})} L^{2/3} + {\mathcal O}(1),
\ee
which for $L=2n$ can be written as
\be
\langle m \rangle_{d^*}^{\rm c} \approx \frac{\Gamma(\frac{2d^*+3}{3})}{\Gamma(\frac{2d^*+2}{3})} (2n)^{2/3} + {\mathcal O}(1),
\ee

\item{$z>1$}

In this case, and when $d$ is of order $n$, the leading asymptotics of $\gierP(2n,d;z)$ will be of the form $p(n)z^n$, where $p(n)$ is a polynomial in $n$. This means that $\theta \gierP(2n,d;z)$ is of the same order as $n\gierP(2n,d;z)$ and (\ref{eq:ohyp}) reduces in leading order to
\begin{equation}
(\theta^3 - 4n^2\theta)\gierP(2n,d;z) = 4z (\theta^3 -
n^2\theta)\gierP(2n,d;z). \label{eq:hypeq>1} 
\end{equation}
Using (\ref{eq:dens}) one can derive the following equation for
$\langle m\rangle_{d^*}^{\rm c}$,
\be
4n^2(z-1) \langle m\rangle_{d^*}^{\rm c} = 
(4z-1) \left( \theta^2 \langle m\rangle_{d^*} +3\langle m\rangle_{d^*}^{\rm c}\theta
\langle m\rangle_{d^*}^{\rm c} + (\langle m\rangle_{d^*}^{\rm c})^3 \right),
\ee
which in leading order when $\langle m\rangle_{d^*}^{\rm c} \sim n$ reduces to
$(4z-1) (\langle m\rangle_{d^*}^{\rm c})^2 = 4n^2(z-1)$ and thus
\begin{equation}
\langle m\rangle_{d^*}^{\rm c} \approx \sqrt{\frac{z-1}{4z-1}}\; L+ {\mathcal O}(1). 
\end{equation}

\end{itemize}
\bigskip

The scaling behaviour near the phase transition at $z=1$ is governed by a single
\index{cross-over exponent}
exponent, the cross-over exponent $\phi$ \cite{Bind83}. On general grounds one expects,
\be
\langle m\rangle_{d^*}^{\rm c} \sim \left\{
\renewcommand{\arraystretch}{1.4}
\begin{array}{ll}
L(z-1)^{1/\phi-1} & (z>1)\\
L^\phi & (z=1)\\
(1-z)^{-1} & (z<1)
\end{array}\right.
\label{eq:avkscalcrit}
.
\ee
Indeed, we find such scaling behaviour for $\langle m\rangle_{d^*}^{\rm c}$ with $\phi=2/3$.

\subsubsection{Scaling function}
\begin{figure}[ht]
\centerline{
\begin{picture}(220,175)
\put(0,10){\includegraphics[width=220pt]{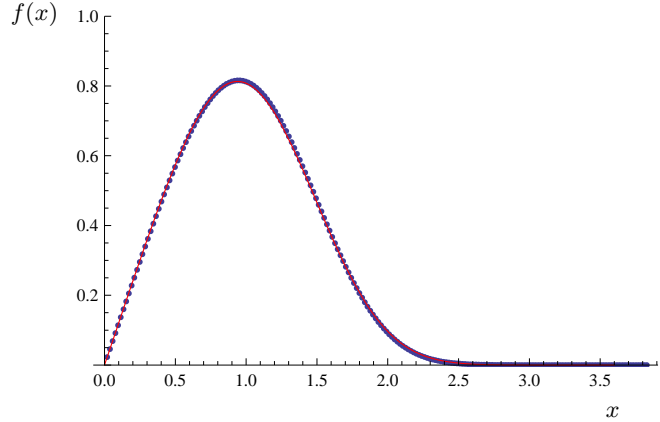}}
\put(-25,150){{$f(x)$}}
\put(200,0){{$x$}}
\end{picture}}
\caption{The scaling function $f(x)$ defined in \eqref{scalfun2} compared to a numerical evaluation (dots) of \eqref{eq:scalfun} for $L=2n=600$.} 
\label{fig:scalfun}
\end{figure}

\index{scaling function}
In \cite{GierNPR04} an analysis has been carried out to obtain the nest scaling function for $L=2n$ and 
\index{nest!scaling function}
$d=n-1$, i.e. $d^*=0$. Following Polyakov \cite{Pol70}, we expect the following scaling form of the nest distribution function at the critical point,
\be
\frac{P(2n,n-1,m)}{S(2n,n-1)} \sim \frac{1}{\langle 1+m\rangle_0} f\left( \frac{1+m}{\langle 1+m\rangle_0}\right)\qquad (n\rightarrow \infty),
\label{eq:scalfun}
\ee
where $\langle 1+m\rangle_0 = 1+ \langle m\rangle_0^{\rm c}$.
The large $x$ behaviour of $f(x)$ is related to the exponent $\phi$ \cite{Cloiz90},
\be
\lim_{x\rightarrow\infty} f(x) \sim x^s \giere^{-a x^\delta},\qquad \delta=\frac{1}{1-\phi},
\ee
where $a$ and $s$ are constants. The behaviour of $f(x)$ for small $x$ is related to the large $n$ behaviour of the probability $P(2n,n-1,m)/S(2n,n-1)$,
\be
\lim_{x\rightarrow 0} f(x) = b x^\vartheta\quad \Rightarrow \quad
b = \lim_{m\rightarrow 0} \lim_{n\rightarrow \infty} \left(1+\langle m\rangle_0\right)^{1+\vartheta} \frac{P(2n,n-1,m)}{S(2n,n-1)},
\label{eq:smallx}
\ee
from which we find 
\be
\vartheta=1,\qquad b = \frac{3}{\Gamma(2/3)^3}.
\ee
\index{scaling function}
Assuming that the \textit{full} scaling function is of the form $x^\vartheta \giere^{-a x^{\delta}}$, for all values of $x$, and using the normalisation condition
\be
\int_0^{\infty} f(x)\gierd x =1,
\ee
we find that
\be
f(x) = b x\, \giere^{-b x^3/3}.
\label{scalfun2}
\ee
In Figure~\ref{fig:scalfun} we compare the scaling function \eqref{scalfun2} with a numerical evaluation of \eqref{eq:scalfun} for $n=300$. When $\delta^*>0$, the value of $\phi$, and hence that of $\delta$, is not changed, but it follows from \eqref{eq:smallx} that the value of the exponent $\vartheta$ changes to
\be
\vartheta=1+2d^*.
\ee
The full scaling function is not known in this case.

\subsection{Half turn symmetry}
\index{fully packed loops!half-turn symmetric}
The following analysis closely follows that of the previous section. We are
interested in the asymptotics as $n\rightarrow \infty$ of the average
number of nests defined in (\ref{eq:dens}). This can be inferred
from the hypergeometric equation for $\gierP(2n;z)$, 
\index{hypergeometric equation}
\begin{equation}
\theta(\theta+1+2n)(\theta+1-2n)\gierP(2n;z) =
4z(\theta+3/2)(\theta+1-n)(\theta+1+n) \gierP(2n;z). \label{eq:chyp}
\end{equation}
Again we discriminate the cases $z<1$, $z=1$ and $z>1$ and remind the reader that $L=2n$.
\begin{itemize}
\item{$z<1$}

In this case, up to an overall constant prefactor, $\gierP(2n;z)$ will grow as a polynomial in $n$ and,
neglecting lower order terms, (\ref{eq:chyp}) reduces to,
\begin{equation}
\theta \gierP(2n;z) = z(\theta +3/2)\gierP(2n;z),
\end{equation}
so that we find $(1-z)\gierP'(2n;z) = \frac32 \gierP(2n;z)$ and thus 
\begin{equation}
\langle 1+m\rangle \approx \frac{2+z}{2(1-z)}+ {\mathcal O}(1).
\end{equation}
\item{$z=1$}

For this case, and exact expression was given for $\langle m+1\rangle$ in Conjecture~\ref{con:avnestHT}. Asymptotically we find
\begin{equation}
\langle 1+m\rangle = n\prod_{j=1}^{n-1} \frac{3j+1}{3j+2} \approx \frac{
\Gamma(5/6)}{\sqrt{\pi}}\; L^{2/3} +\mathcal{O}(1).
\end{equation}
\item{$z>1$}

Here,up to an overall constant prefactor, $\gierP(2n;z)$ will grow as $p(n)z^n$ where $p(n)$ is a polynomial in
$n$. This means that $\theta \gierP(2n;z)$ will be of the same order as $n \gierP(2n;z)$ and (\ref{eq:chyp}) reduces in leading order to
\begin{equation}
(\theta^3 - 4n^2\theta)\gierP(2n;z) = 4z (\theta^3 - n^2\theta)\gierP(2n;z),
\end{equation}
which is the same as (\ref{eq:hypeq>1}). We thus find that
\begin{equation}
\langle 1+m\rangle \approx \sqrt{\frac{z-1}{4z-1}}\; L +\mathcal{O}(1). 
\end{equation}

\end{itemize}
For half-turn symmetric FPL diagrams we find the same cross-over exponent $\phi=2/3$ as for horizontally symmetric FPL diagrams.

\section{Conclusion}
We have described a model of tightly packed, nested polygons on the square lattice. We hope that the study of such tightly packed polygons is relevant to other polygon models described in this book. The advantage of the model described in this chapter is that many exact results can be obtained, even on finite geometries, due to its relation with the exactly solvable six-vertex and O($n=1$) lattice models. In particular, the statistical mechanical partition function can be obtained rigorously on finite square patches of the square lattice. The free energy can then be obtained analytically and in the thermodynamic limit. The fully packed loop model undergoes a well known bulk order-disorder phase transition as a function of an anisotropy parameter associated to the straight segments of the polygon boundary.

We have furthermore shown that it is possible to obtain closed form expressions for partition functions of important subsets of fully packed loop configurations. Two examples of such subsets are horizontally symmetric fully packed loop models of depth $d$, and half-turn symmetric  fully packed loop models of depth $d$. These closed form expressions have been obtained experimentally, and remain conjectures at the time of this writing. In addition, using hypergeometric summation identities, we were able to compute the average number of polygon nests at the boundary in closed form. Asymptotic analyses allowed us to study a boundary phase transition as a function of the nest fugacity, and we obtained a crossover exponent $\phi=2/3$. At criticality, we derive a scaling form for the nest distribution function which displays an unusual non-Gaussian cubic exponential behaviour.

To conclude it should be said that while some of the exact results presented in this chapter are more than what one would hope for from a physicist's perspective, where numerical techniques are often all what is available, they are just the starting point for a mathematician. Although great progress has been made in recent years in understanding fully packed loop models, proving conjectures such as the nest distribution function and many related combinatorial results, remains a fascinating and completely open problem. It is for reasons such as these that polygon models in all shapes and sizes will continue to inspire future research.

\newcommand\arxiv[1]{{\tt arXiv:#1}}


\end{document}